# Revealing the ultra-sensitive calorimetric properties of superconducting magic-angle twisted bilayer graphene

G. Di Battista[1], P. Seifert[1], K. Watanabe[2], T. Taniguchi[3], K.C. Fong[4], A. Principi[5] and D. K. Efetov[1]*

1. ICFO - Institut de Ciencies Fotoniques, The Barcelona Institute of Science and Technology, Castelldefels, Barcelona, 08860, Spain
2. Research Center for Functional Materials, National Institute for Materials Science, 1-1 Namiki, Tsukuba 305-0044, Japan
3. International Center for Materials Nanoarchitectonics, National Institute for Materials Science, 1-1 Namiki, Tsukuba 305-0044, Japan
4. Quantum Engineering and Computing Group, Raytheon BBN Technologies, Cambridge, Massachusetts 02138, United States
5. Department of Physics and Astronomy, The University of Manchester, M13 9PL Manchester, United Kingdom

*E-mail : dmitri.efetov@icfo.eu

**The allegedly unconventional superconducting phase of magic-angle twisted bilayer graphene (MATBG)[1] has been predicted to possess extraordinary thermal properties, as it is formed from a highly diluted electron ensemble with both a record-low carrier density $n \sim 10^{11}$ cm$^{-2}$ and electronic heat capacity $C_e < 100$ $k_B$. While these attributes position MATBG as a ground-breaking material platform for revolutionary calorimetric applications[2], these properties have so far not been experimentally shown. Here we reveal the ultra-sensitive calorimetric properties of a superconducting MATBG device, by monitoring its temperature dependent critical current $I_c$ under continuous laser heating with a wavelength of $\lambda = 1550$nm. From the bolometric effect, we are able to extract the temperature dependence of the electronic thermal conductance $G_{th}$, which remarkably has a non-zero value $G_{th} = 0.19$ pW/K at 35mK and in the low temperature limit is consistent with a power law dependence, as expected for nodal superconductors. Photo-voltage measurements on this non-optimized device reveal a peak responsivity of $S = 5.8 \times 10^7$ V/W when the device is biased close to $I_c$, with a noise-equivalent power of $NEP = 5.5 \times 10^{-16}$ WHz$^{-1/2}$. Analysis of the intrinsic performance shows that a theoretically achievable limit is defined by thermal fluctuations and can be as low as $NEP_{TEF} < 10^{-20}$ WHz$^{-1/2}$, with operation speeds as fast as $\tau \sim 500$ ns. This establishes superconducting MATBG as a revolutionizing active material for ultra-sensitive photon-detection applications, which could enable currently unavailable technologies such as THz photon-number-resolving single-photon-detectors.**

To date the most sensitive detectors for electro-magnetic radiation are based on superconducting materials and exploit local photo-induced heating across their strongly temperature dependent superconducting transition[3–6]. To maximize the temperature increase due to absorbed radiation, the key requirements for such materials are an ultra-low electronic heat capacity $C_e$[7], which is typically achieved by using nano-structured thin films[6], as well as a good thermal isolation and ultra-low thermal conductance $G_{th}$ to its surroundings[8]. Possessing both of these attributes, graphene has recently attracted formidable attention[9–13]. While not an intrinsic superconductor, it can be proximitized by superconducting electrodes to form Josephson junctions, and as such was successfully used as a GHz bolometer[14,15] and mid-IR single-photon detector[16,17]. Furthermore the recently discovered moiré material, magic angle twisted bilayer graphene (MATBG) was shown to have an intrinsic, gate tunable superconducting phase with a record-low carrier density $n < 10^{11}$ cm$^{-2}$ and electronic heat capacity $C_e < 100$ $k_B$ [1,2], where $k_B$ is the Boltzmann constant. Since both quantities are by several orders of magnitude lower than for any other superconductor, this establishes MATBG as an extremely promising superconducting material for next generation calorimetric applications[2].

The signature superconducting properties of MATBG are shown in Fig. 1a and b. The typical device consists of a van der Waals stack of graphite/hBN/MATBG/hBN where the MATBG has a twist-angle of 1.10°±0.03° and is encapsulated into insulating hBN hexagonal boron nitride layers (see device image in the inset of Fig. 1a). By applying a gate voltage $V_g$ to the metallic graphite layer we can electrostatically tune the carrier concentration in the MATBG sheet $n$. Fig. 1a shows measurements of the longitudinal resistance $R_{xx}$ vs. electronic temperature $T_e$ at optimal doping, which reveal a sharp superconducting transition with a critical temperature of $T_c \sim 2.1$ K (defined by the temperature at which $R_{xx}$ equals to 50% of the normal state resistance) and a high peak value $dR_{xx}/dT > 8$ kΩ/K. The superconducting region is dome shaped in the $n$-$T$ phase space and lies in close proximity to a correlated insulating state, which occurs at a filling of electrons per moiré unit cell of $\nu = -2$ (Fig. 1b (top)). The correlated insulator leads to a Fermi level reset, as is evidenced from measurements of the Hall density $n_H$ vs. carrier density $n$ (Fig. 1b (bottom)), which dramatically lowers the density of the free carriers. For optimal doping of the superconducting state we extract an ultra-low free carrier density of only $n_H = -1.96 \times 10^{11}$ cm$^{-2}$, which directly translates into a record low electronic heat capacity per area $C_e/A \sim 10^2\, k_B\, \mu m^{-2}$, as is calculated in Fig. 1c for the same parameters.

Here we test the thermal and calorimetric properties of MATBG electrons in the superconducting regime. For this purpose, we illuminate (spot-diameter ~ 1.9mm) the device with laser light at telecom wavelength of $\lambda = 1550$nm, and monitor its 4-terminal transport properties, as shown schematically in Fig. 1d. We use the temperature dependence of the critical current $I_c$ to calibrate the electronic temperature $T_e$ and this thermometry scheme to accurately monitor the heating effects induced by light illumination (see SI for more details). This allows us to work out the calorimetric sensitivity of the superconducting state of MATBG and to determine its electronic thermal conductance $G_{th}$. In addition, we study the performance of MATBG as a photodetector and estimate its noise equivalent power *NEP*.

While the exact processes which govern light-matter interactions in MATBG are not yet experimentally studied, we assume that for our experimental parameters these are very similar to those in AB bilayer graphene sheets[18], as the schematics in Fig. 1e shows. For the near infrared wave-lengths applied here, photons do not couple to phonons in the system and are absorbed only by the electrons with an absorption percentage of ~ 4.6% [19], by exciting optically electron-hole pairs (e-h) in the higher energy dispersive bands. These assumptions are justified for MATBG as the energy of the incident photons of ~ 0.8 eV is orders of magnitude larger than the width of the flat-bands[20] ~ 10 meV and the size of superconducting gap ~1meV[21]. It is in principle also possible that photons are directly absorbed by Cooper pairs, which are broken and excited into higher lying bands. However as the $k$-space of the flat-bands does not extend far away from the Dirac points[22], this results in a vanishing joint density of states for vertical transitions from and to the flat-bands. Moreover, the wavelength of the photons is an order of magnitude smaller than the size of the moiré unit cell ~ 13nm[1,20] , making the effect of the super-lattice on this process negligible.

The absorbed photon energy is then subsequently transferred into heat, where the excited e-h pairs relax down the bands and eventually thermalize with the electrons in the flat-bands. The transferred heat in this process raises the electronic temperature of the flat-bands $T_e$ above the device temperature, which is given by the lattice and the leads and is well thermalized with the bath temperature $T_b$. Since the temperature of the electrons is elevated by $\Delta T_e = T_e - T_b$ above the bath temperature, they dissipate heat into their colder surroundings. This cooling process is defined by the electronic thermal conductance $G_{th}$, which we assume to be predominantly governed by electron-electron (Wiedemann-Franz) interactions for temperatures below 1K[23].

To study the thermal properties of the MATBG device in the superconducting state we first calibrate its electronic temperature $T_e$. This can be achieved by transport measurements of the critical current of the superconducting state $I_c$, which is a monotonic function of its electronic temperature $T_e$. To accomplish this, we perform non-linear resistance measurements $dV_{xx}/dI$ vs. source-drain current $I_{dc}$, where $I_c$ is extracted from the maxima of the $dV_{xx}/dI$ ($I_{dc}$) traces. By heating up the bath temperature of the cryostat $T_b$, which is in equilibrium with the electronic temperature of the MATBG $T_e$, we can define a direct correspondence of $I_c$ ($T_e$), as is shown in Fig. 2a (top) and Fig. 2b (left).

Next, we repeat a similar calibration but now by heating the electrons with laser light, while keeping the bath temperature constant $T_b$ = 35mK. As discussed earlier, incident radiation only couples to electrons and heats these above the bath temperature $T_e > T_b$, which remains unchanged under the minute laser power. To estimate the exact power that is absorbed by the electrons $P_L$, we carefully calibrate the power that is incident on the device (see SI), and adjust the absorption to 4.6%, as is expected for graphene bilayers[19]. Fig. 2a (bottom) shows the corresponding $dV_{xx}/dI$ ($I_{dc}$) vs. $P_L$ measurements and Fig. 2b (right) shows the extracted $I_c$ ($P_L$). Overall these measurements show striking similarities to the previous measurements vs. bath temperature, and confirm our assumptions that the absorption of radiation primarily induces a bolometric effect via heating the electron gas.

The precise measurement of $I_c$ as a function of both, the electronic temperature $I_c$ ($T_e$) and absorbed laser power $I_c$ ($P_L$), allows us to infer the bolometric effect of the MATBG electrons. In order to obtain a smooth calibration between $T_e$ and $P_L$ we first fit the $I_c$ ($T_e$) data in Fig. 2b (left) with the empirical relation: $I_c(T_e) = I_c(T_e = 0)[1 - (T_e/T_c)^4]^{3/2}$, which is expected from BCS theory of a superconductor[24](see SI for more information). Inserting the $I_c$ ($P_L$) measurements from Fig. 2b (right) into this relation allows us to obtain the dependence of the electronic temperature on the absorbed radiation $T_e$ ($P_L$).

We now apply these methods and measure $T_e$ vs. $P_L$ as a function of $T_b$ between 35 mK and 810 mK, the highest temperature at which we can accurately determine $I_c$ (Fig. 2c). For low heating power and small $\Delta T_e$ the linear response regime holds[25], which allows us to define the electronic thermal conductance $G_{th}(T_b) = P_L/\Delta T_e$, a quantity which is a direct function of the bath temperature $T_b$, as shown in Fig. 3a. For $T_b$ = 35mK we extract a $G_{th}$ = 0.19 pW/K. This extremely low number proves the very good thermal isolation of the electrons in MATBG. Considering the theoretical estimation of heat capacity (see SI) of Fig. 1c, we can now also predict the thermal relaxation time of MATBG electrons, which for low power excitations is given by the ratio[25] $\tau_{th} = C_e/G_{th}$. The resulting $\tau_{th} \sim$ 500 ns are shorter than that for most state-of-the-art transition-edge sensors[5,8] and could enable faster operation speeds[2,15].

The temperature dependence of the electronic thermal conductance in the superconducting state has been famously used to determine the symmetry of the superconducting gap[26–28]. In s-wave superconductors, owing to their isotropic superconducting gap, thermal conductance follows an exponential activation behavior, where thermal excitations are efficiently blocked for temperatures $T_e << T_c$. This is in contrast to nodal p- and d-wave superconductors, where the thermal conductance follows a power law temperature dependence[27], and the nodes of the superconducting gap allow thermal excitations even for $T_e << T_c$. Since the order parameter of the superconducting state of MATBG has not been measured so far, we attempt to shed light on its symmetry, by theoretically modelling the extracted $G_{th}$ ($T_b$) behavior. As phonons are typically frozen below $T_b$ < 1K, we assume that electron-phonon scattering can be neglected[23]. Hence the thermal conductance below this temperature is dominated by the Wiedemann-Franz law, where only electrons that are thermally excited above the SC gap conduct heat, while Cooper pairs do not. We model using the one-dimensional heat transfer equation for the local temperature $T_e(x)$, as is in detailed described in the SI.

In Fig. 3a we show the modeled $G_{th}$ ($T_b$) for the cases of an isotropic s-wave and for a p- or d-wave superconducting gap, which are best fits for the experimental data. Here the size of the superconducting gap $\Delta_0$ is the only fitting parameter, while we fix the experimentally obtained parameters $T_c$ = 2.1K and normal state resistivity $\rho$ = 20kΩ. We note that the two cases yield a significantly different thermal conductance for the limit of $T_e << T_c$. Especially for $T_b$ < 0.8 K, the thermal conductance for an isotropic superconductor becomes exponentially small, while that of a nodal superconductor decays with a much slower power law dependence. The later provides a much better fit of the experimental data, which also has non-zero values even at the smallest temperatures. Here, the best fit for the isotropic case gives a $\Delta_0$ ~ 0.42 ± 0.06 meV, while for the nodal case we obtain a $\Delta_0$ ~ 1.0 ± 0.2 meV, which is in very good agreement with recent experimental reports on gap size measurements of MATBG of $\Delta_0$ ~ 1 meV[21]. These results suggest that the obtained thermal conductance is rather consistent with a nodal p- or d-wave symmetry, than with an isotropic s-wave symmetry, showing overall good qualitative and quantitative agreement with theory, which reassures the validity of the assumptions and findings.

We now estimate the sensitivity of the MATBG superconductor to irradiation with light. For this sake we measure the differential photo-voltage $V_{ph}$ vs. dc current $I_{dc}$ as a function of absorbed laser power $P_L$, as is shown in Fig. 4a (see Methods). We find that the photo-voltage monotonically increases as $I_{dc}$ is enhanced and has a pronounced maximum when the critical current is reached $I_{dc} \sim I_c$ (Fig. 4a). This is explained by a photo-induced temperature increase when the device is biased close to $I_c$, which drives the MATBG into the resistive state and generates a voltage peak, similarly to previously reported superconducting detectors[29]. Fig. 4b shows the power dependence of the voltage peaks $V_{ph}^{max} = V_{ph}(I_c)$ taken at $T_b$ = 35mK. Its value is almost constant for the lowest $P_L$, however as $P_L$ is increased it follows a linear power dependence, which indicates that the detector operates in a linear response regime. From the slope we can extract a high responsivity $S$ = 5.8 x $10^7$ V/W. We define the noise level $V_N$ as the standard deviation of $V_{ph}^{max}$ at low powers divided by the square root of the equivalent noise bandwidth (see SI), where we find a value of $V_N \sim 10^{-8}$ VHz$^{-1/2}$. With these experimental findings we can now also estimate the as measured noise equivalent power of the device $NEP = V_N/S$ = 5.5 x $10^{-16}$ WHz$^{-1/2}$. This number is already lower than previously proposed graphene bolometers[9,12,13] but still 3 orders of magnitude higher than the best transition-edge sensors[8,30] reported in the literature. However, neither the measurement circuit nor device were yet optimized for detector applications.

To work out the intrinsic performance limits of a potential MATBG detector we estimate also the theoretically achievable *NEP*s (see Fig. 4c). First, we compute the *NEP* which is limited by thermal Johnson noise $NEP_{Johnson} = \sqrt{4k_BRT_b}/S$, where $R$ is the detector resistance, dominated by the contact resistance (typically ~ 10 kΩ). Here for 35mK we obtain $NEP_{Johnson}$ = 2.4 x $10^{-18}$ WHz$^{-1/2}$. Using the experimentally extracted thermal conductance we can further estimate the intrinsic *NEP* limit, which is defined by the thermal fluctuations of the bath. For 35mK we get $NEP_{TEF} = \sqrt{4G_{th}k_BT_b{}^2} = 1.1 \times 10^{-19}$ WHz$^{-1/2}$. These estimates allow to immediately propose avenues to optimize the detector design. Correspondingly, scaling the device dimensions to reduce $C_e$ and $NEP_{TEF}$ and optimizing the operation scheme[31] as in other transition-edge sensors[3,4] can allow to reach lower *NEP*s. Proper device engineering can also considerably reduce the contact resistance[32] and highly suppress the Johnson noise. These adjustments could allow to reach $NEP_{TEF} < 10^{-20}$ WHz$^{-1/2}$ , which is comparable to state-of-the-art transition-edge sensors[8,30]. Specifically, as has been theoretically shown before[2], MATBG can be used as the active material for THz and sub-THz superconducting bolometers as well as photon-number-resolving single photon detectors of low-energy photons[2].

In conclusion, the coexistence of record-small carrier density, ultra-low electronic heat capacity and thermal conductance, and a highly temperature-sensitive, gate tunable resistance in superconducting MATBG makes it a highly promising material platform for sensitive photodetection applications. We estimate that MATBG superconductors could reach *NEP*s $< 10^{-20}$WHz$^{-1/2}$ when operated in the thermal limit and could so enable calorimeters, bolometers, transition edge sensors and single photon detectors with unprecedented sensitivity from the visible to THz and even GHz frequencies, and with thermal response times as low as 500 ns, which is close to the speed of current superconducting qubits readout systems[15,33,34].

Acknowledgements:

D.K.E. acknowledges support from the Ministry of Economy and Competitiveness of Spain through the “Severo Ochoa” program for Centres of Excellence in R&D (SE5-0522), Fundació Privada Cellex, Fundació Privada Mir-Puig, the Generalitat de Catalunya through the CERCA program, funding from the European Research Council (ERC) under the European Union’s Horizon 2020 research and innovation programme (grant agreement No. 852927)". G.D.B. acknowledges support from the “Presidencia de la Agencia Estatal de Investigación” (Ref. PRE2019-088487). K.W. and T.T. acknowledge support from the Elemental Strategy Initiative conducted by the MEXT, Japan (Grant Number JPMXP0112101001) and JSPS KAKENHI (Grant Numbers 19H05790, 20H00354 and 21H05233).

Author contributions:

D.K.E and G.D.B. conceived and designed the experiments; G.D.B. fabricated the devices and performed the measurements; G.D.B., D.K.E. and A. P. analyzed the data; A. P. performed the theoretical modeling; T.T. and K.W. contributed materials; D.K.E., P.S. and K. C. F. supported the experiments: G.D.B., D.K.E. and A.P. wrote the paper.

**Supplementary Information** is available for this paper.

**Correspondence and requests for materials** should be addressed to D.K.E.

**Competing interest.** Authors claim no competing interest.

## Methods

Device fabrication.

The fabrication procedure for MATBG devices has two main steps: the first one in which the layers are exfoliated and assembled to form a van der Waals heterostructure and the second one in which the structure is patterned and the electric contacts are made. The stack was assembled using the cut and stack technique. A hBN flake was picked up at 100 °C with a stamp of propylene carbonate (PC)/polydimethyl siloxane (PDMS) mounted on a glass slide. The first half of graphene (pre-cut using an AFM tip) was picked up with the hBN flake. Subsequently, the second graphene layer was rotated to an angle of 1.1° and picked up with the hBN/graphene stack. The heterostructure was then fully encapsulated with another hBN layer. A graphite layer was picked up in the last step to act as local gate. The stack was then deposited melting the PC film at 180 °C on an O2-plasma-cleaned Si/SiO2 chip. At this point the structure was patterned by Electron Beam Lithography to form a Hall bar and etched with CHF3/O2 mixture to expose graphene edges which were then contacted by metal leads Cr/Au (5/50 nm).

Transport and photovoltage measurements.

The measurements were carried out in a dilution refrigerator (BlueFors-SD250) at a base temperature of 35 mK. Standard low-frequency lock-in techniques (Stanford Research SR860) were used to measure the longitudinal $R_{xx}$ and transverse $R_{xy}$ resistance in 4-probe configuration with an excitation current of 10 nA at a frequency of 17 Hz. For the measurements of the differential voltage $dV_{xx}/dI$, a dc signal (generated using Keithley 2400 source meter) passed through a 1/10 voltage divider in series with a 1-MΩ resistor and was combined with an ac excitation current of 2 nA. The differential voltage signal was then amplified with a SR560 low-noise voltage preamplifier and measured at the same frequency of 17 Hz with the Stanford Research SR860 lock-in. All the current-bias measurements were performed applying a dc voltage (Keithley 2400) signal through a 1/10 divider and a 1-MΩ resistor. In order to shine light on the device a single frequency 1550nm laser was brought via a single-mode optical fiber into the dilution refrigerator. A collimator was mounted few centimetres on top of the device to shine collimated light with a spot diameter of 1.9 mm. The device was carefully centered in the middle of the spot. The incident laser power was adjusted using a variable optical attenuator (JGR OA5 l). For the differential photovoltage experiments, the laser emission was modulated with a sinusoidal profile. The photovoltage was then amplified with a SR560 room-temperature voltage preamplifier and measured with the lock-in referenced at the same frequency of the modulation.

Twist angle extraction.

To extract the twist angle between the two graphene layers, we analyze the Hall measurements at low magnetic field (Extended Data Figure 1). When we apply a gate voltage $V_g$ to the graphite, the gate-induced carrier density $n$ varies according to the formula $n = V_g C_g/e$ where $C_g$ is the gate capacitance of the bottom hBN layer and $e$ the elementary charge of the electron. Since near charge neutrality the Hall charge carrier density ($n_H = -B/(eR_{xy})$) equals the gate-induced carrier density $n$, performing a linear fit we can accurately extract $C_g$. From the position of band insulators (Extended Data Figure 1) we can estimate the charge carrier density corresponding to the fully filled superlattice unit cell $n_s$ which directly relates to the twist angle $\theta$: $n_S = 8\theta^2/\sqrt{3}a^2$, where $a$=0.246 nm is the interatom distance in single layer graphene. The twist angle we determine is $\theta$=1.10°±0.03°.

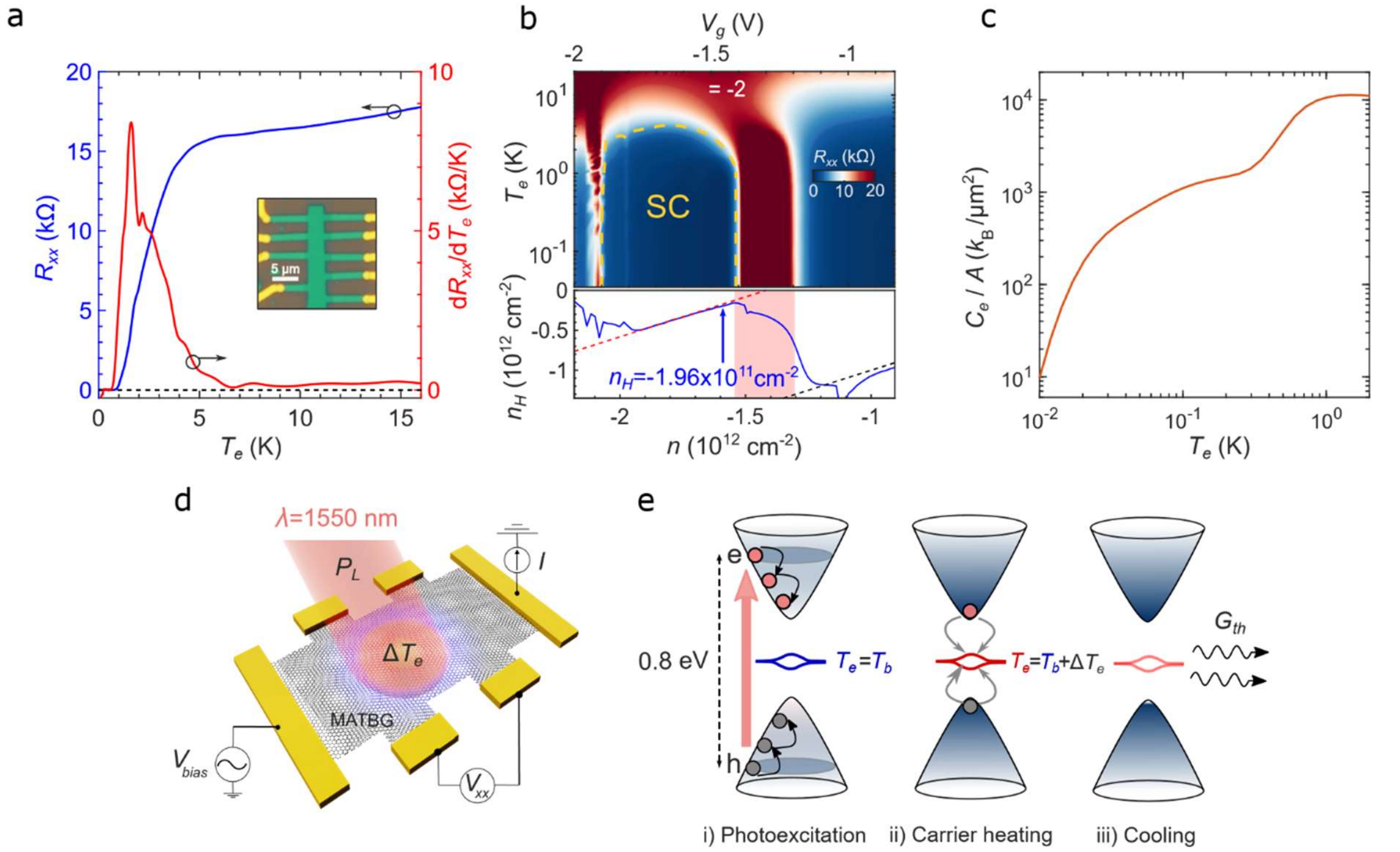

Fig. 1**. Ultra-sensitive calorimetry based on MATBG superconductors.** Fig.1 (a) and (b) show transport measurements of a typical superconducting MATBG device with twist-angle of 1.10° (**a**) Longitudinal resistance $R_{xx}$ and numerical derivative d$R_{xx}$ /d$T_e$ vs electron temperature $T_e$ at a fixed carrier density $n_H$ = -1.96x$10^{11}$ cm$^{-2}$. Inset shows image of the measured device. (**b**) Top panel: $R_{xx}$ vs gate voltage $V_g$ and temperature $T_e$, shows a dome shaped superconducting region (yellow dashed line) which is flanking a correlated insulating state at a filling factor $v$ = -2 (white dashed line). Bottom panel: low-field Hall density $n_H$ extracted at $B$ = 300 mT as a function of gate-induced carrier density $n$. At $v$ = -2 we observe a Fermi level reset, which sets the free carrier density of the corresponding superconducting state to an ultra-low value of $n_H$ = -1.96x$10^{11}$ cm$^{-2}$. (**c**) Calculation of the electronic heat capacity per area of MATBG for the same carrier density as in (a) and (b), showing ultra-low values of $C_e/A \sim 10^2$ $k_B$ μm$^{-2}$ at 35 mK. (**d**) Schematic of the bolometric measurements, which employ four-terminal transport measurements under uniform light illumination with a wavelength of λ = 1550nm. (**e**) Schematics of light-matter interactions in MATBG. The absorbed photons generate electron-hole pairs in the high order bands which thermalize in the flat bands. The cooling process is ruled by the electronic thermal conductance $G_{th}$.

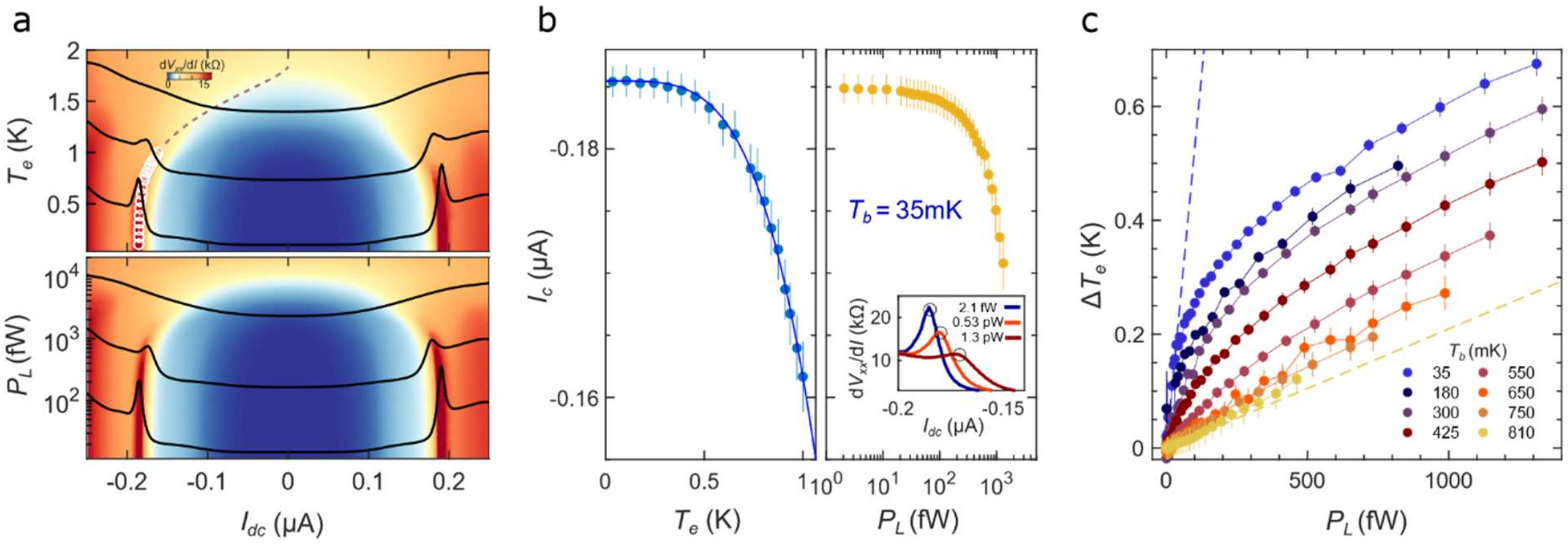

Fig. 2**. Calorimeter calibration and bolometric response.** (**a**) Top panel: color plot of the longitudinal d$V_{xx}$/d$I$ vs dc current bias $I_{dc}$ and temperature $T_e$. The dashed line is the fit of the critical current with the empirical expression for a superconductor $I_c(T)= I_c(T=0)[1-(T_e/T_c)^4]^{3/2}$. Bottom panel: color plot of the longitudinal d$V_{xx}$/d$I$ vs $I_{dc}$ and laser power $P_L$. In both panels the white dots are the extracted values of critical current $I_c$, while the black lines are line-cuts at 3 different values of temperature (top panel) and power (bottom panel). (**b**) Critical current $I_c$ versus $T_e$ (right panel) and $P_L$ (right panel). The blue line is the fit according to the empirical formula used in (a). The inset in the right panel displays the method we use to extract the critical current (black dots) from the peak of d$V_{xx}$/d$I$ at different powers. (**c**) Variation of electronic temperature $\Delta T_e$ as a function of laser power $P_L$ for different bath temperatures ($T_b$) from 35mK to 810mK. The blue (yellow) dashed line is the linear fit of the data at 35mK (810mK) from which we extract $G_{th}$ assuming the linear regime approximation.

a

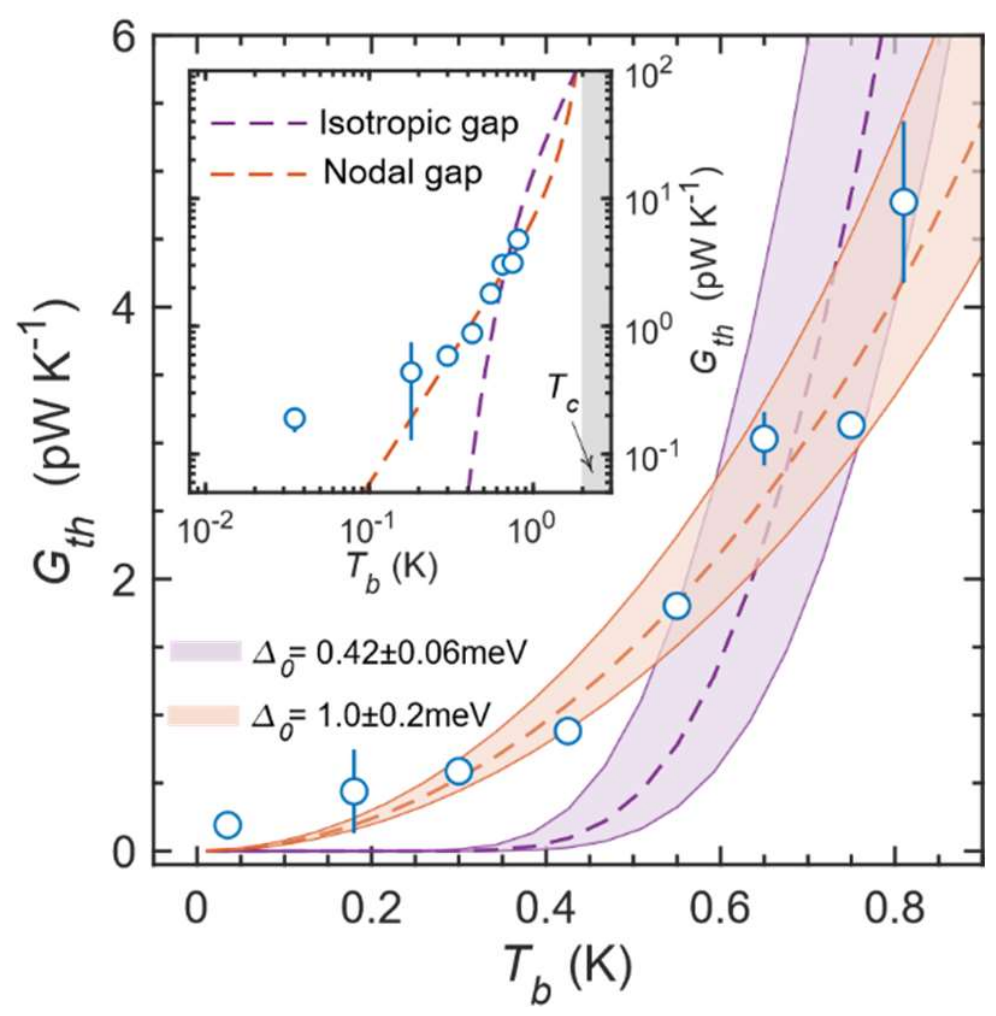

Fig. 3**. Thermal conductance.** (**a**) Measured thermal conductance $G_{th}$ at different bath temperatures ($T_b$). The dashed lines are the best fits to the experimental data of the modeled $G_{th}$ assuming Wiedemann-Franz law, for the case of an isotropic s-wave (violet) and a nodal superconducting gap (orange). This model is obtained fixing $T_c = 2.1$K, the normal state resistivity $\rho = 20$kΩ and using the size of the superconducting gap $\Delta_0$ as the only fitting parameter. The shaded regions represent the modeled $G_{th}$ in the range $\Delta_0 \sim 0.42 \pm 0.06$ meV (for s-wave) and $\Delta_0 \sim 1.0 \pm 0.2$ meV (for nodal superconducting gap). As inset, the same data in logarithmic scale up to $T_c$, denoted as the grey-shaded region.

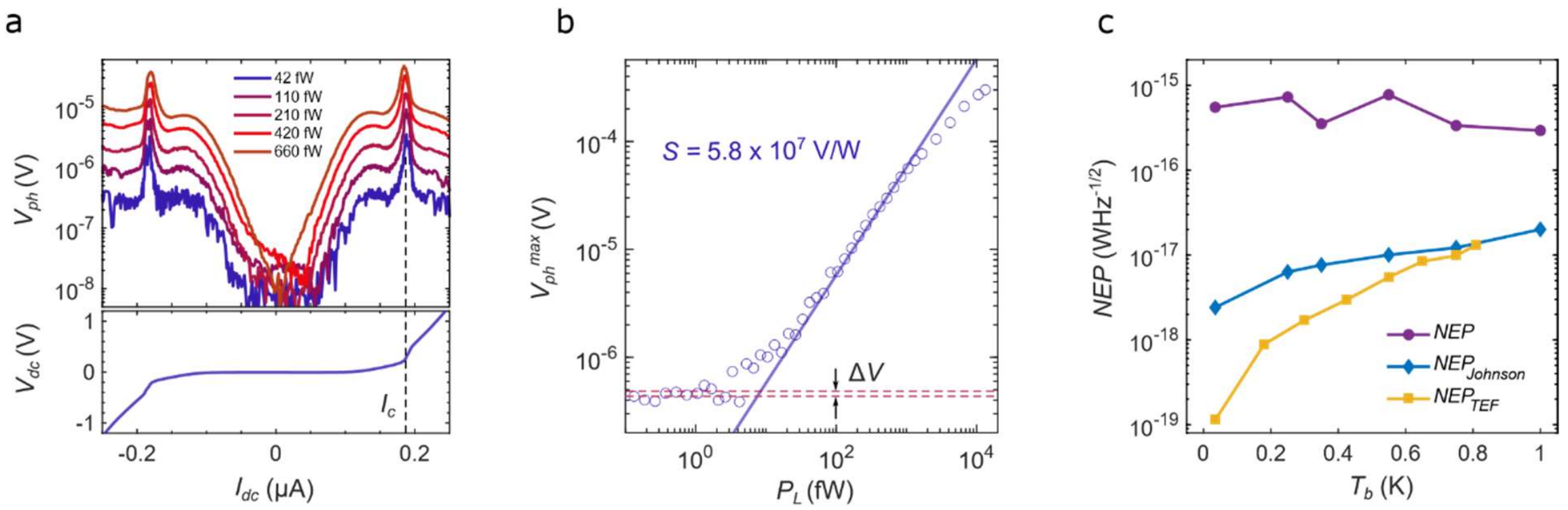

Fig. 4**. Detector performance and *NEP*. (a)** Top panel: differential photovoltage $V_{ph}$ as function of dc current bias ($I_{dc}$), measured at different laser powers. Bottom panel: Current-voltage characteristic *I/V* as function of dc current bias ($I_{dc}$). $V_{ph}$ shows pronounced peaks at the critical current. **(b)** Extracted photovoltage peaks $V_{ph}^{max}$ as a function of $P_L$. From the linear fit (blue line) we extract the responsivity *S*. The red dashed lines mark the standard deviation of $V_{ph}^{max}$ at low powers ($\Delta V$). **(c)** Noise-equivalent power *NEP* for different operating temperatures. In violet we show the as measured *NEP*, in blue the theoretically predicted Johnson noise limited $NEP_{Johnson}$ and in yellow the ultimate limit imposed by the thermal fluctuations $NEP_{TEF}$.

# Supplementary information: Revealing the ultra-sensitive calorimetric properties of superconducting magic-angle twisted bilayer graphene

G. Di Battista[1], P. Seifert[1], K. Watanabe[2], T. Taniguchi[3], K.C. Fong[4], A. Principi[5] and D. K. Efetov[1]*

1. ICFO - Institut de Ciencies Fotoniques, The Barcelona Institute of Science and Technology, Castelldefels, Barcelona, 08860, Spain
2. Research Center for Functional Materials, National Institute for Materials Science, 1-1 Namiki, Tsukuba 305-0044, Japan
3. International Center for Materials Nanoarchitectonics, National Institute for Materials Science, 1-1 Namiki, Tsukuba 305-0044, Japan
4. Quantum Engineering and Computing Group, Raytheon BBN Technologies, Cambridge, Massachusetts 02138, United States
5. Department of Physics and Astronomy, The University of Manchester, M13 9PL Manchester, United Kingdom

*E-mail: dmitri.efetov@icfo.eu

**Table of Contents**

## A. Table of the measured device

| | | | |
|---|---|---|---|
| MATBG area dimensions (μm) | length=4.1, width=3.7 | | |
| Twist angle $\theta$ | 1.10°±0.03 | | |
| $V_g$ (V) | -1.455 | -1.480 | -1.495 |
| $C_g/e$ ($cm^{-2}/V$) | 1.09 $10^{12}$ | | |
| Carrier density $n$ ($10^{12}$ $cm^{-2}$) | -1.59 | -1.61 | -1.63 |
| Hall carrier density $n_H$ ($10^{12}$ $cm^{-2}$) | -0.196 | -0.212 | -0.231 |
| $G_{th}$ (pW $K^{-1}$) at $T_b$=35mK | 0.19±0.04 | 0.19±0.05 | 0.12±0.04 |
| Responsivity $S$ (V/W) at $T_b$=35mK | 5.8 $10^7$ | 3.0 $10^7$ | 1.7 $10^7$ |
| Thermal fluctuation limited $NEP$ ($NEP_{TEF}$) ($WHz^{-1/2}$) at $T_b$=35mK | 1.1 $10^{-19}$ | 1.1 $10^{-19}$ | 9.0 $10^{-20}$ |
| $NEP_{TEF}/\sqrt{Area}$ ($WHz^{-1/2}\mu m^{-1}$) at $T_b$=35mK | 2.9 $10^{-20}$ | 2.9 $10^{-20}$ | 2.3 $10^{-20}$ |
| Measured $NEP$ ($WHz^{-1/2}$) at $T_b$=35mK | 5.5 $10^{-16}$ | 6.4 $10^{-16}$ | 7.1 $10^{-16}$ |
| $T_c$(K) (50% normal state resistance) | 2.1 | 2.6 | 2.8 |

## B. Extended transport data

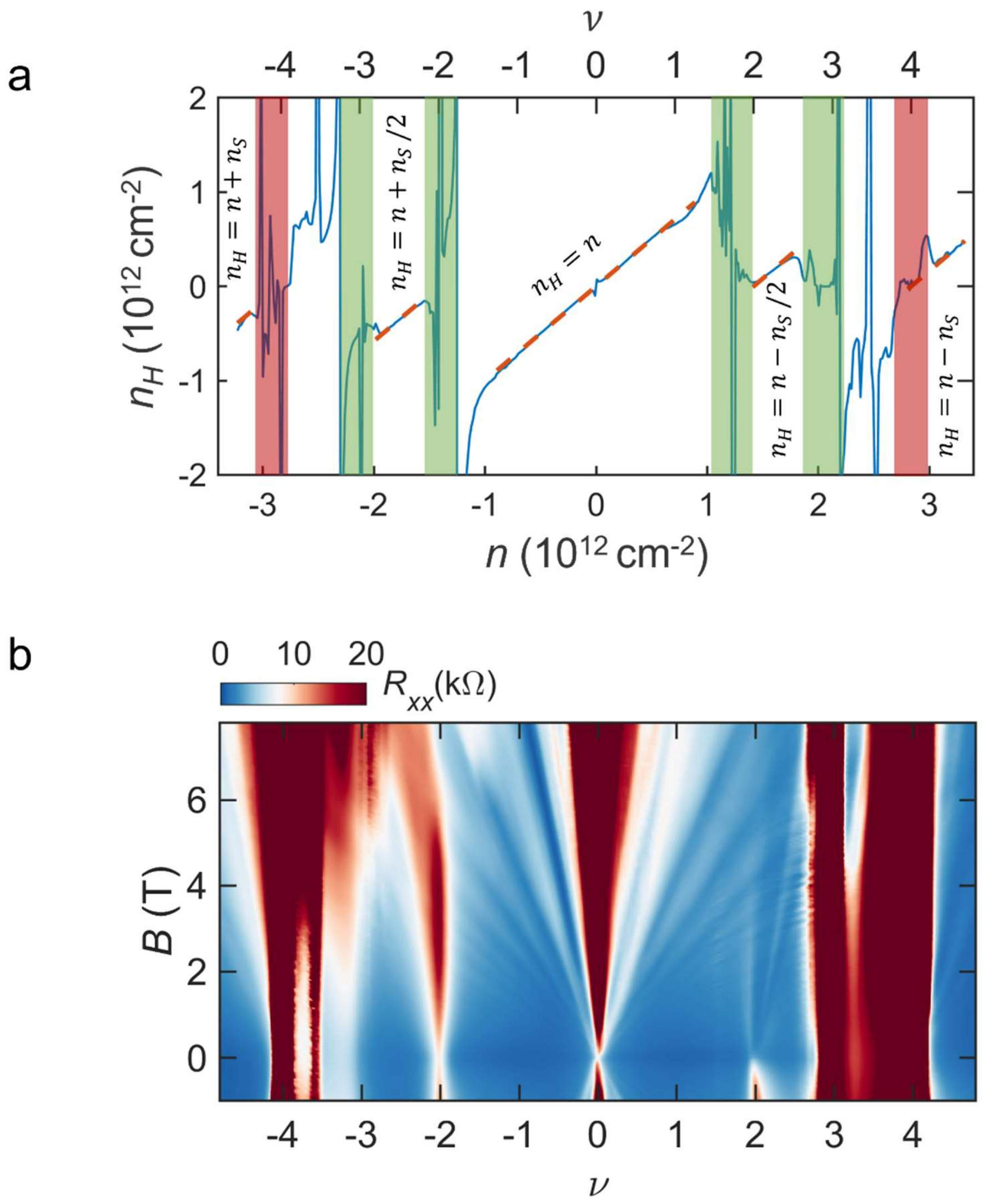

Extended data Figure 1.| **Hall density measurements and Landau fan. (a)** The light-blue line trace shows the Hall carrier density $n_H$ versus the gate-induced carrier density $n$ taken at 0.3 T and 35 mK. The light green (light red) stripes indicate the position of correlated insulating states (band insulating states), at which we observe clear signatures of Hall density resets. The orange dashed lines show that the Hall carrier density follows $n_H = n$ near charge neutrality, $n_H = n \pm n_s/2$ beyond the half-filling correlated states and $n_H = n \pm n_s$ after the band insulators, where $n_s$ is the carrier density corresponding to the fully filled superlattice unit cell. **(b)** Landau fan diagram at 1.5 K. The pronounced quantum oscillations demonstrate the quality and the cleanliness of the MATBG device.

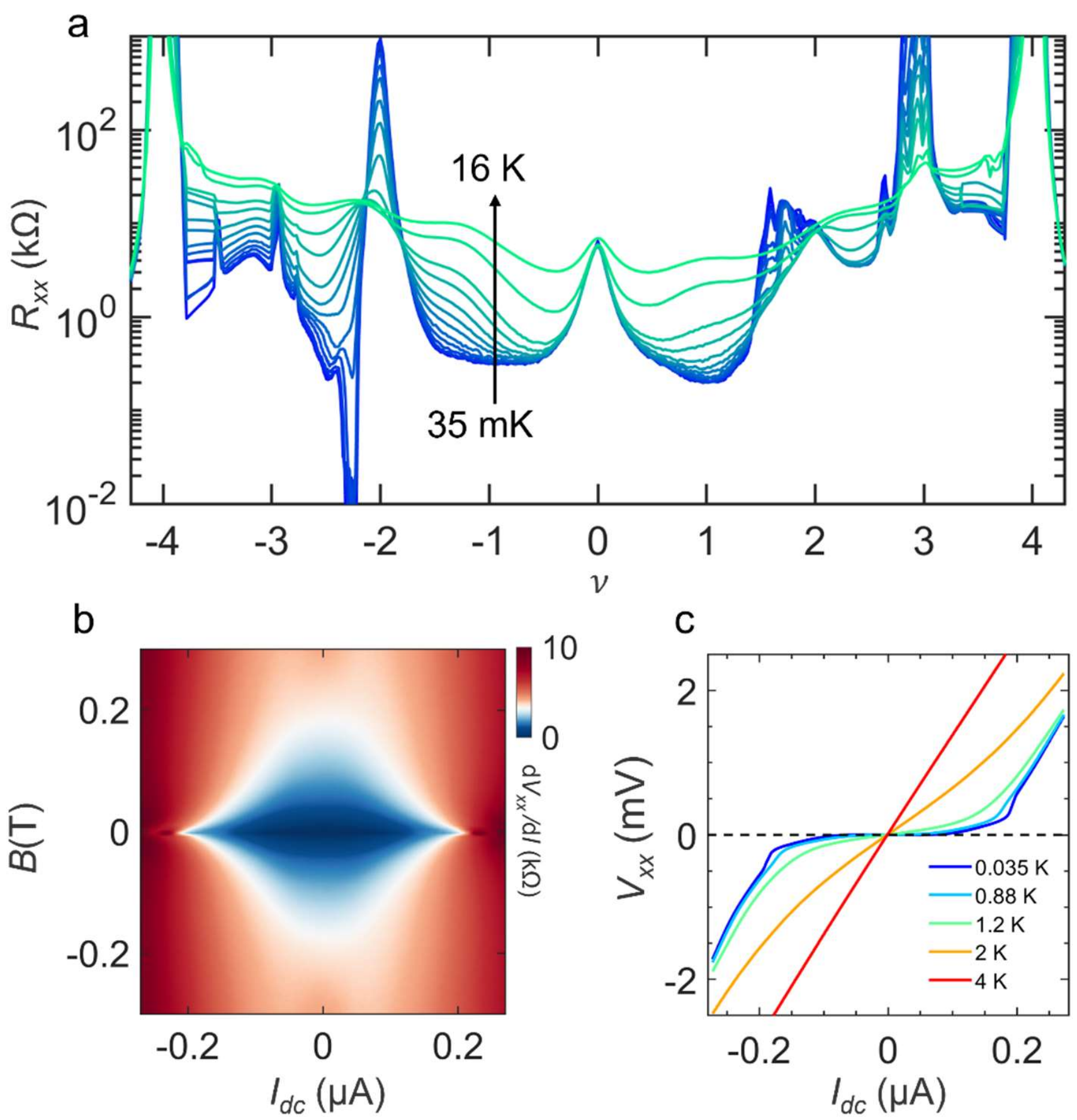

Extended data Figure 2. | **Additional data on the superconducting phase. (a)** Longitudinal resistance $R_{\mathrm{xx}}$ versus filling factor $\nu$ for different temperatures ranging from 35 mK (blue) to 16 K (green). The zero-resistive state beyond $\nu$=-2 is the superconductor studied in the experiment. **(b)** Differential resistance $dV_{xx}/dI$ as a function of current bias $I_{dc}$ and magnetic field $B$ for $\nu$=-2.24. The ac excitation current used for this measurement is $I_{ac}$ = 2 nA. The superconducting state is smeared out by magnetic field. **(c)** Current-voltage characteristics $I/V$ at various temperatures for $\nu$=-2.24. At the lowest temperature of 35 mK, the critical current is around 0.18 μA.

## C. Calorimeter calibration formula

To derive the formula used to fit $I_c$ vs $T_e$ we consider that in the BCS theory the upper critical current density $J_c$ for the existence of a superconducting phase depends on the product between the superconducting gap $\Delta$ and the local density of superconducting electrons[1] $N_s$: $J_c \sim N_s\Delta$. Using the definition of the order parameter, that $N_s \sim \Delta^2$ we get $J_c \sim N_s^{3/2}$. Since to fit our data we need a formula which is valid at all temperatures and not just close to $T_c$, we use the empirical relation[1] for the superconducting electron density $N_s(T)=N_s(0)[1-(T_e/T_c)^4]$. From this assumption we obtain the expression:

$$I_c(T_e) = a\left[1-\left(\frac{T_e}{b}\right)^4\right]^{3/2} \tag{C.1}$$

Where $a$ and $b$ are two fitting parameters. $a$ is an estimation of $I_c(T_e=0)$ while $b$ is an estimation of $T_c$. This qualitative argument provides an empirical formula that reproduces our experimental data and allows us to calibrate the calorimeter, finding an analytical expression to relate temperature $T_e$ and power $P_L$:

$$T_e(P_L) = b\left[1-\left(\frac{I_c(P_L)}{a}\right)^{2/3}\right]^{1/4} \tag{C.2}$$

When calibrating the calorimeter at different bath temperatures $T_b$, to avoid any artifacts which may arise from an offset of $I_c$ between different measurements we write the critical current as $I_c(P_L)=\Delta I_c(P_L)+ I_c(T_b)$. $I_c(T_b)$ is the critical current predicted from the fit in (*C.1*) at a certain bath temperature $T_b$. Since $I_c$ is constant for the lowest powers, $\Delta I_c(P_L)$ is defined as the variation of critical current from the average at low powers given by $\Delta I_c(P_L)= I_c(P_L)-\langle I_c\rangle_{lowP}$. Thus, the final calibration expression reads:

$$T_e(P_L) = b\left[1-\left(\frac{\Delta I_c(P_L)+I_c(T_b)}{a}\right)^{2/3}\right]^{1/4} \tag{C.3}$$

Similarly, since $T_e$ is constant for the lowest powers to obtain the variation of electronic temperature $\Delta T_e$ we substract the average at low powers: $\Delta T_e(P_L)= T_e(P_L) -\langle T_e\rangle_{lowP}$.

## D. Calculation of the laser power on the device

As detailed in the methods section, in our setup we couple a telecom laser which emits an output power $P_{out}$ with a single-mode optical fiber designed for 1550-nm transmission. To calculate the effective power illuminating the MATBG device we consider a Gaussian beam profile. In this approximation, the intensity profile $I$ as a function of the distance from the beam center $r$ and distance away from the end of the fiber $z$ reads[2]:

$$I(r,z) = I_0 \left(\frac{w_0}{w(z)}\right)^2 e^{-2(r/w(z))^2} \tag{D.1}$$

Where $w_0$ is the beam radius at the end of the fiber and $I_0 = 2P_{out}/(\pi\, w_0^2)$ the total irradiance coming out of the laser source imposing the Gaussian normalization condition. $w(z)$ is the value of the radius at a distance $z$ from the fiber given[2] by $w(z) = w_0\sqrt{1+(z/z_R)^2}$ where $z_R$ is the Rayleigh range. In our case, since we use a collimator, we can consider $w(z) \simeq w_0$=0.95 mm. Considering our device a rectangle with sizes $l_1$ and $l_2$ placed at the center of the beam we estimate the effective power ($P_L$) absorbed by the MATBG as:

$$P_L = \eta\, \alpha\, T\, I_0 \int_{-l_1/2}^{+l_1/2} dx \int_{-l_2/2}^{+l_2/2} dy\; e^{-2\frac{x^2+y^2}{w_0^2}} \tag{D.2}$$

Where $\alpha$ is the estimated absorption of the MATBG sheet ($\alpha$=4.6%) for 1550-nm photons, $T$ the effective transmission of the fiber and $\eta$ the variable attenuation we use to control the power incident on the device.

## E. Extended data on Photovoltage and *NEP* extraction

We estimate the measured noise-equivalent power ($NEP_{meas}$) as the ratio between the experimentally measured noise level $V_N$ and the extracted responsivity $S$:

$$NEP_{meas} = \frac{V_N}{S} \tag{E.1}$$

For each $P_L$ we find the bias at which the detector shows the highest photo-response. In Extended data Figure 3a we plot the photovoltage maxima ($V_{ph}^{max}$) extracted for each power and perform a linear fit (black line), which gives the responsivity (in V/W).

The noise level is defined by the fluctuations of $V_{ph}^{max}$ at low powers, where the detector response is too low to be detected. We estimate these fluctuations calculating the standard deviation of $V_{ph}^{max}$ at low powers:

$$\Delta V = \sqrt{< V_{ph}^{max}(lowP_L)^2 > - < V_{ph}^{max}(lowP_L) >^2} \tag{E.2}$$

$V_N$ then reads:

$$V_N = \frac{\Delta V}{\sqrt{ENBW}} = \frac{\Delta V}{\sqrt{8 * T_{const}}} \tag{E.3}$$

Where ENBW is the equivalent noise bandwidth[3] which depends upon the time constant ($T_{const}$ =300 ms) and the filter roll off (12 dB) used in the lock-in during the experiment. We repeat this measurement at different $T_b$ to probe the detector performances even at temperatures higher than base temperature of the cryostat.

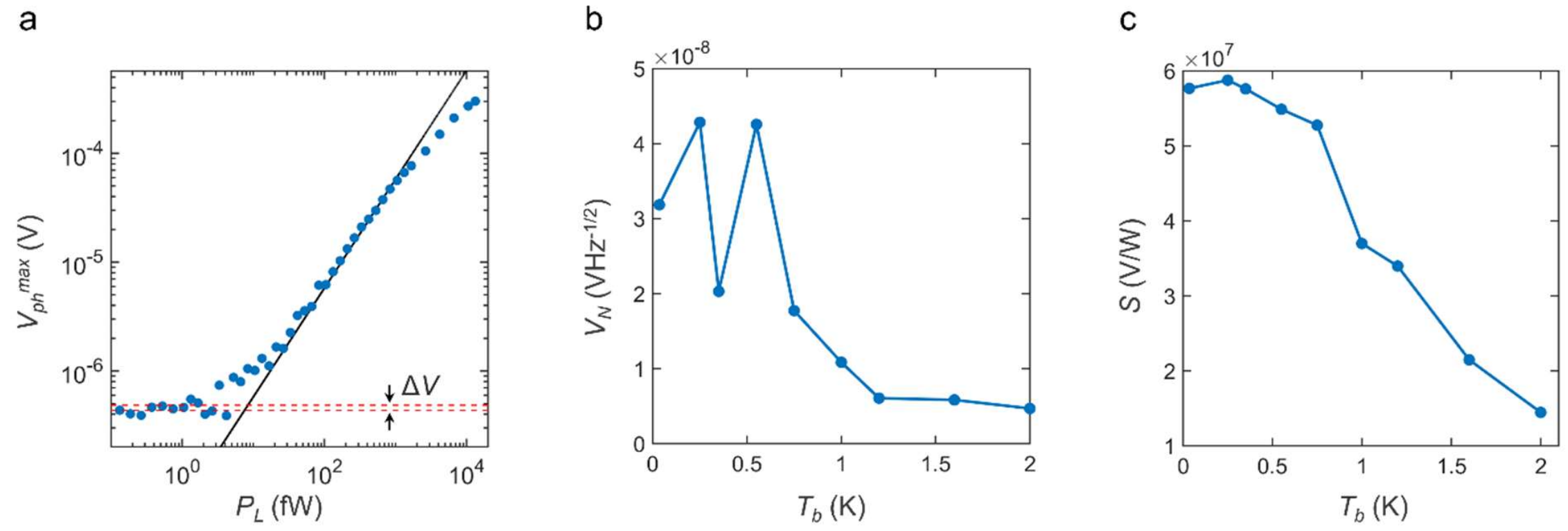

Extended data Figure 3. | **Responsivity and noise level at different temperatures. (a)** Photovoltage maxima $V_{ph}^{max}$ extracted at each laser power $P_L$. The black line is a linear fit from which we extract the responsivity. The red dashed lines indicate the standard deviation of the $V_{ph}^{max}$ at low powers. **(b)** Noise level measured at different bath temperatures. **(c)** Extracted responsivity at different bath temperatures.

In Extended data Figure 4a we show the measured photovoltage traces vs bias current at different laser powers for $T_b$ = 35 mK. When the device is biased close to the critical current $I_{dc}$~$I_c$ it becomes very instable and shows peaks even at very low powers (blue traces in Extended data Figure 4a) or without illumination. This results in a relatively high noise floor as shown in Extended data Figure 3b, which limits the detector performances. On the contrary for $I_{dc}$~0, there are not such peaks and the noise floor measured is 2 orders of magnitude lower, as in Extended data Figure 4b ($V_N(I_{dc}=0) \sim 10^{-10}$ VHz$^{-1/2}$). This observation confirms that the measurement circuit is not optimized and that a MATBG can reach better performances than the one reported in the main text. In analogy to what it is usually done in transition-edge sensors[4], a possible strategy to reduce this noise close to the transition could be to operate the detector in a voltage-bias scheme so that the bolometer is maintained at a fixed temperature through negative electro-thermal feedback.

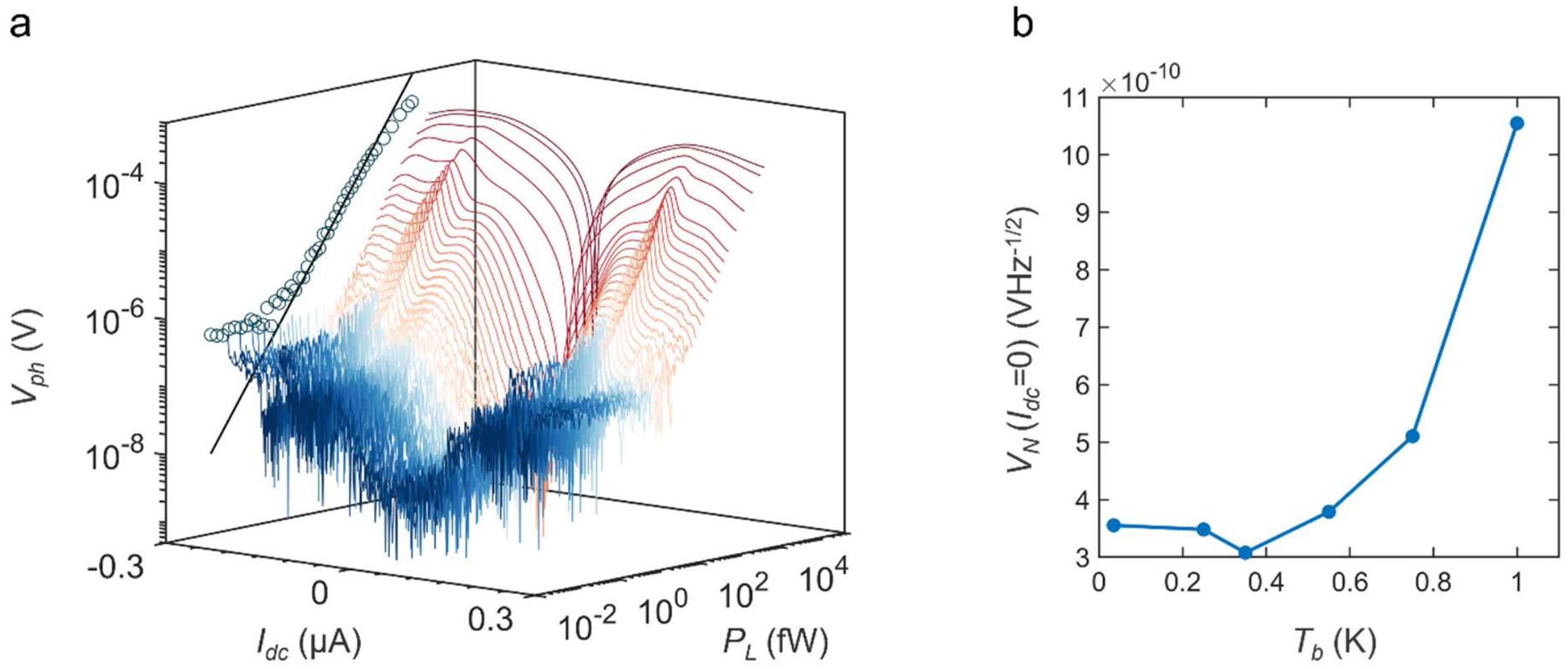

Extended data Figure 4.| **Extended data on photovoltage and noise level. (a)** Photovoltage traces vs bias current at different powers. In the projection of the left panel the extracted maxima $V_{ph}^{max}$ which are fitted to extract the responsivity. **(b)** Noise level measured at different bath temperatures when the device is biased at $I_{dc}=0$. $V_N$ is obtained according to the equation (E.3).

## F. Gate tunability of the superconducting state

Another novel feature of MATBG is that the physical properties of the superconducting state namely the carrier density and the sharpness of the transition can be tuned applying an external voltage. Extended data Figure 5a reports the $R_{xx}$ vs $T_e$ and the d$R_{xx}$/d$T_e$ for 5 different gate voltages $V_g$ in the superconducting dome. The dots in yellow and red are the maxima of d$R_{xx}$/d$T_e$ and the extracted $T_c$ (temperature at which $R_{xx}$ equals to 50% of the normal state resistance) respectively. $T_c$ ranges from ~2K to 3.5K and peaks at the center of the dome. Strikingly, the transition reaches the steepest point ~10kΩ/K close to the insulating state at $V_g$=-1.455V (corresponding to $\nu$=-2) and decreases sharply moving away from it. That doping is also the one with the lowest carrier density, meaning that the MATBG calorimeter operates at its best as close as possible to the correlated insulating state. As a result, measuring the responsivity of the device at different gate voltages (Extended data Figure 5b) we find that $V_g$=-1.455V is the doping which shows the highest responsivity.

The gate-tunability of the device enables the possibility to finely control the detector performances just by applying an external voltage. This represents an important novelty compared with traditional 3D transition-edge sensors based on non-gate-tunable metals for which the superconducting properties are intrinsic in the material.

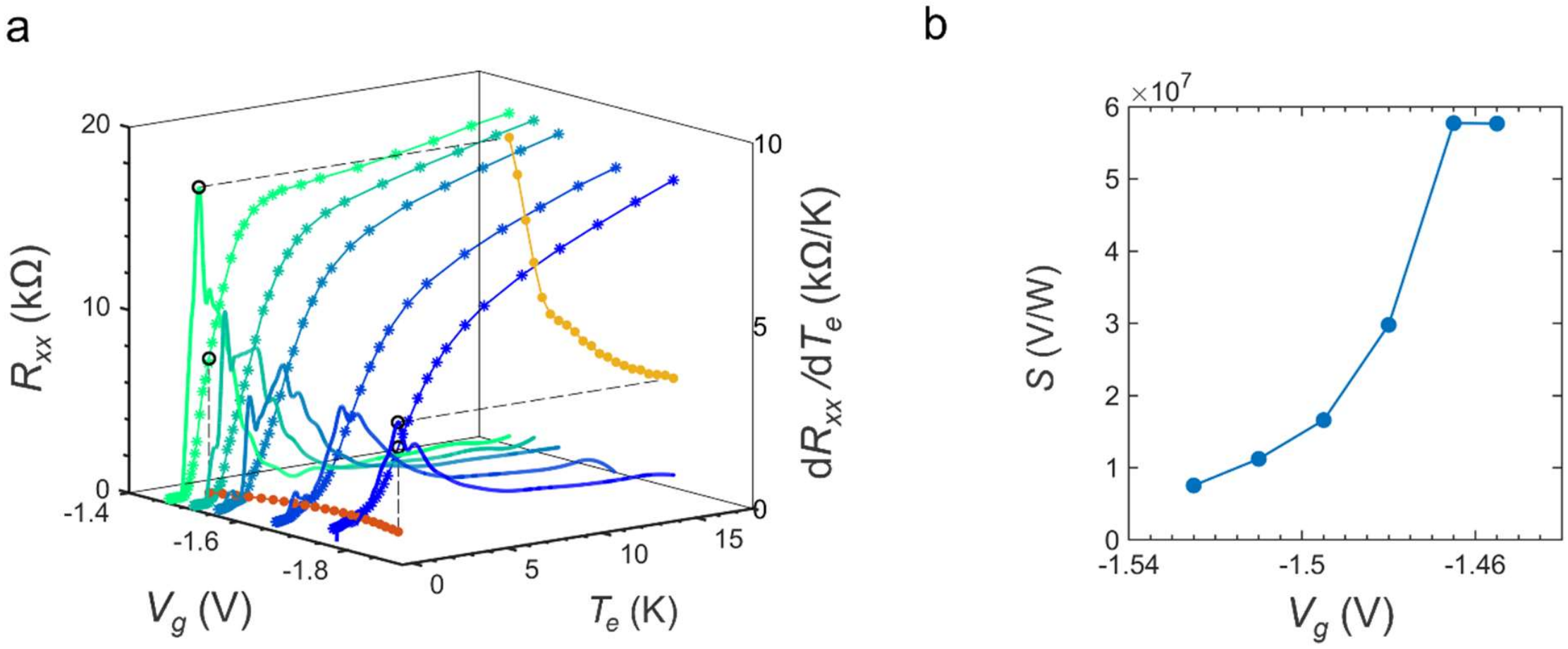

Extended data Figure 5. | **Gate tunability of the superconducting state. (a)** Longitudinal resistance versus temperature $R_{xx}$ vs $T_e$ (left-axis) and d$R_{xx}$/d$T_e$ vs $T_e$ (right-axis) for different gate voltages in the superconducting dome. The yellow data projected on the right are the extracted d$R_{xx}$/d$T_e$ for each gate voltage while the red data projected on the bottom are the critical temperatures (defined as the temperature at which $R_{xx}$ equals to 50% of the normal state resistance). **(b)** Extracted responsivity $S$ at different gate voltages.

## G. Theoretical Calculations of electronic heat capacity

To determine the electronic heat capacity $C_e$, we start from the kinetic equation for the distribution function of electrons with momentum $\boldsymbol{k}$ and in band $\lambda$, $f_{\boldsymbol{k},\lambda}$, in the absence of external fields and particle flow, *i.e.*

$$\partial_t f_{\boldsymbol{k},\lambda} = I[f_{\boldsymbol{k},\lambda}], \tag{G.1}$$

where $I[f_{\boldsymbol{k},\lambda}]$ is the collision integral (for the present calculation, its specific form is not important). We now multiply the left-hand side of Eq. (G.1) by $\varepsilon_{\boldsymbol{k},\lambda} - \mu$, where $\varepsilon_{\boldsymbol{k},\lambda}$ is the electron band energy and $\mu$ is the chemical potential, and we integrate it over $\boldsymbol{k}$ and sum over $\lambda$. We then identify the resul with $C_e\ \partial_t T_e$, where the electronic heat capacity reads:

$$C_e = \sum_{\boldsymbol{k},\lambda} (\varepsilon_{\boldsymbol{k},\lambda} - \mu) \left(-\frac{\partial f_{\boldsymbol{k},\lambda}}{\partial \varepsilon_{\boldsymbol{k},\lambda}}\right)\left(\frac{\varepsilon_{\boldsymbol{k},\lambda} - \mu}{T_e} - \frac{\partial \mu}{\partial T_e}\right). \tag{G.2}$$

Here,

$$\frac{\partial \mu}{\partial T_e} = \frac{\sum_{\boldsymbol{k},\lambda}\left(-\frac{\partial f_{\boldsymbol{k},\lambda}}{\partial \varepsilon_{\boldsymbol{k},\lambda}}\right)\left(\frac{\varepsilon_{\boldsymbol{k},\lambda} - \mu}{T_e}\right)}{\sum_{\boldsymbol{k},\lambda}\left(-\frac{\partial f_{\boldsymbol{k},\lambda}}{\partial \varepsilon_{\boldsymbol{k},\lambda}}\right)}. \tag{G.3}$$

This equation is obtained by assuming the carrier density to be independent of the electron temperature $T_e$ and fixed, e.g., by an external gate.

## H. Theoretical Calculations of thermal conductance

We consider a gas of quasiparticles and quasiholes in a superconducting channel of length $L$ and width $W$, heated by an external laser. We assume that the boundaries of the systems $x = 0, L$ are kept at the initial temperature of the bath $T_b$, and that quasiparticles can only transport and dissipate heat there. The thermal conductance of the channel, $G_{th} = \frac{P_L}{T_e - T_b}$, is defined as the coefficient which relates the total input power $P_L$, which is assumed to be constant throughout the channel, to the rise in average temperature of electrons, $T_e - T_b$.
To find this coefficient we solve the one-dimensional heat equation[5,6] for the local electron temperature $T_e(x)$ (0<x< $L$):

$$\kappa \partial_x^2 T_e(x) + \frac{P_L}{WL} = 0 \tag{H.1}$$

where $\kappa$ is the quasiparticle (Wiedemann-Franz) thermal conductivity, which is determined in the following. We assume that $|T_e(x) - T_b|/T_b \ll 1$, *i.e.* the input power is sufficiently small as to not perturb the superconducting state too much. Solving Eq.(H.1) with the boundary conditions $T_e(0) = T_b$ and $T_e(L) = T_b$, and averaging over the length of the channel we get:

$$T_e - T_b \equiv \frac{1}{L}\int_0^L dx\,[T_e(x) - T_b] = -\frac{P_L}{\kappa W L^2}\int_0^L dx \frac{x(x-L)}{2} = \frac{P_L}{\kappa W L^2}\frac{L^3}{12} \tag{H.2}$$

and therefore the channel thermal conductance is identified as:

$$G_{th} = \frac{12\kappa W}{L} \tag{H.3}$$

## i. Model of the superconductor

Since most of the physics at low temperatures occurs in the vicinity of the Fermi surface, we will describe electrons in twisted bilayer graphene as an electron gas whose density is coupled to a short-ranged quenched scalar disorder. The Hamiltonian is:

$$\mathcal{H} = \frac{1}{2}\sum_{\boldsymbol{k},\sigma,\sigma'} \left(c^\dagger_{\boldsymbol{k},\sigma}, c_{-\boldsymbol{k},\sigma'}\right)\begin{pmatrix} \xi_{\boldsymbol{k}} & \Delta_{\boldsymbol{k},\sigma,\sigma'} \\ \Delta_{\boldsymbol{k},\sigma,\sigma'} & -\xi_{-\boldsymbol{k}} \end{pmatrix}\begin{pmatrix} c_{\boldsymbol{k},\sigma} \\ c^\dagger_{-\boldsymbol{k},\sigma'} \end{pmatrix} + \sum_{\boldsymbol{k},\boldsymbol{k}',\sigma} u_{\boldsymbol{k}-\boldsymbol{k}'} c^\dagger_{\boldsymbol{k}',\sigma} c_{\boldsymbol{k},\sigma} \tag{H.4}$$

In these equations, $c^\dagger_{\boldsymbol{k},\sigma}$ ($c_{\boldsymbol{k},\sigma'}$) creates (destroys) an electron of momentum $\hbar\boldsymbol{k}$, spin $\sigma \in \{\uparrow,\downarrow\}$ and energy $\xi_{\boldsymbol{k}} = \hbar v_F^\star(k - k_F)$ ($v_F^\star$ is the Fermi velocity and $k_F$ the Fermi wave vector), while $u_{\boldsymbol{k}}$ is the Fourier transform of the electron-impurity interaction $u(\boldsymbol{r})$. The latter is chosen to be randomly distributed with zero average and short-ranged, *i.e.* $\langle u(\mathbf{r})\rangle = 0$ and $\langle u(\boldsymbol{r})u(\boldsymbol{r}')\rangle = u_0^2\delta(\boldsymbol{r}-\boldsymbol{r}')$. Furthermore, $\Delta_{\boldsymbol{k},\sigma,\sigma'}$ is the superconducting gap. In the s-wave (p-wave) case, $\Delta_{\boldsymbol{k},\sigma,\sigma'}$ is nonzero only if $\sigma' \neq \sigma$ ($\sigma' = \sigma$). We have checked that a superconducting gap with d-wave symmetry produces results effectively undistinguishable from the p-wave case. This in turn implies that the dominant feature is the existence or not of a nodal line in the superconducting order parameter, *i.e.* of a finite in-gap density of states. In what follows we will consider the following fully-gapped and nodal order parameter: $\Delta_{\boldsymbol{k}} = \Delta(T)$ and $\Delta_{\boldsymbol{k}} = \Delta(T)k_x/k$, respectively. These pairings couple quasiparticles with opposite or equal spins, respectively, always at opposite ends of the Fermi surface. In both cases, the gap scales according to the phenomenological formula:

$$\Delta(T) = \Delta_0\sqrt{1-\left(\frac{T}{T_c}\right)^4} \tag{H.5}$$

In our experiment the temperature of the superconducting transition $T_c = 2.1$ K is estimated as the point at which $R_{xx}$ equals 50% of the normal state resistance while the zero-temperature gap $\Delta_0$ is used as a fitting parameter.

In both cases, the bare Hamiltonian is diagonalized with the introduction of (fermionic) quasiparticle operators, $\gamma_{k,\pm}$, which are connected to the bare electron spinors of Eq.(H.4) via the transformation matrix:

$$U_{\boldsymbol{k}} = \frac{1}{\sqrt{2}} \begin{pmatrix} \sqrt{1 + \xi_{\boldsymbol{k}}/\varepsilon_{\boldsymbol{k}}} & -\sqrt{1 - \xi_{\boldsymbol{k}}/\varepsilon_{\boldsymbol{k}}} \\ \sqrt{1 - \xi_{\boldsymbol{k}}/\varepsilon_{\boldsymbol{k}}} & \sqrt{1 + \xi_{\boldsymbol{k}}/\varepsilon_{\boldsymbol{k}}} \end{pmatrix} \tag{H.6}$$

The non-interacting quasiparticles spectrum is $\varepsilon_{k,\lambda} = \lambda \varepsilon_k$ where $\varepsilon_k = \sqrt{\xi_k^2 + \Delta_k^2}$ .
To find the matrix elements of the interaction between quasiparticles and impurities, we transform electron operators to $\gamma_{k,\lambda}$ $(\lambda = \pm)$ with the unitary matrix (H.6). We find:

$$\begin{aligned} \left\langle \left| V_{\boldsymbol{k},\lambda;\boldsymbol{k}',\lambda'} \right|^2 \right\rangle &= n_{\text{imp}} u_0^2 \left| \left[ U_{\boldsymbol{k}'}^\dagger \tau_z U_{\boldsymbol{k}} \right]_{\lambda',\lambda} \right|^2 \\ &\simeq \delta_{\lambda,\lambda'} \frac{n_{\text{imp}} u_0^2}{2} \left( 1 + \frac{\xi_{\boldsymbol{k}'} \xi_{\boldsymbol{k}} - |\Delta_{\boldsymbol{k}'}| |\Delta_{\boldsymbol{k}}|}{\varepsilon_{\boldsymbol{k}'} \varepsilon_{\boldsymbol{k}}} \right) \end{aligned} \tag{H.7}$$

where $n_{\text{imp}}$ is the impurity density and we used that, since the scattering is elastic and conserves energy, only intraband processes limit transport.
We connect the electron-impurity interaction $n_{imp} u_0^2$ to measurable quantities by using that the Drude resistivity of the normal state, within the Born approximation, can be written as:

$$\rho = \frac{h}{e^2} \frac{2 n_{\text{imp}} v_0^2}{(\hbar v_{\text{F}}^\star)^2} \tag{H.8}$$

As we show below, besides being a function of $\Delta(T)$, the thermal conductivity depends only one the value of $\rho$, and not on $n_{\text{imp}} v_0^2$ and $v_{\text{F}}^\star$ separately. Therefore, only two independent measurements are needed to determine the theoretical value of $\kappa$. The typical measured resistivity is $\rho \sim 20\text{k}\Omega$.

## ii. Thermal conductance from kinetic equation

The goal of this section is to derive expressions for the quasiparticle thermal conductivity $\kappa$ to be used in Eq. (H.3). To derive this quantity, we consider the linearized kinetic equation for quasiparticle excitations in the steady state, propagating under the effect of a temperature gradient:

$$\left( -\frac{\partial f_{\boldsymbol{k},\lambda}}{\partial \varepsilon_{\boldsymbol{k},\lambda}} \right) \left( \frac{\varepsilon_{\boldsymbol{k},\lambda}}{T} - \frac{\partial \varepsilon_{\boldsymbol{k},\lambda}}{\partial T} \right) v_{\boldsymbol{k},\lambda} \cdot \nabla_{\boldsymbol{r}} T = I_{\boldsymbol{k},\lambda} \tag{H.9}$$

where we assumed that quasiparticles follow a local-quasi-equilibrium Fermi-Dirac distribution $f_{\boldsymbol{k},\lambda}$ characterized by a single temperature $T(\boldsymbol{r}, t)$. In Eq. (H.9), $v_{\boldsymbol{k},\lambda} = \nabla_{\boldsymbol{k}} \varepsilon_{\boldsymbol{k},\lambda}$ is the quasiparticle velocity, $\lambda = \pm$ labels electron/hole branches and $\varepsilon_{\boldsymbol{k},\lambda} = \lambda \sqrt{\xi_{\boldsymbol{k}}^2 + \Delta_{\boldsymbol{k}}^2}$ is the quasiparticle energy.
Finally, $I_{\boldsymbol{k},\lambda}$ models collisions between quasiparticles and impurities:

$$I_{\boldsymbol{k},\lambda} = N_{\mathrm{f}} \sum_{\lambda'} \int \frac{d^2k'}{(2\pi)^2} \left[ f_{\boldsymbol{k},\lambda}(1 - f_{\boldsymbol{k}',\lambda'}) W_{\boldsymbol{k},\lambda\to\boldsymbol{k}',\lambda'} - f_{\boldsymbol{k}',,\lambda'}(1 - f_{\boldsymbol{k},\lambda}) W_{\boldsymbol{k}',\lambda'\to\boldsymbol{k},\lambda} \right] \tag{H.10}$$

where $N_{\mathrm{f}} = 4$ is the number of spin-valley fermion flavors and the transition probability, calculated within the Fermi golden rule, reads:

$$W_{\boldsymbol{k},\lambda\to\boldsymbol{k}',\lambda'} = \frac{2\pi}{\hbar} \left\langle \left| V_{\boldsymbol{k},\lambda;\boldsymbol{k}',\lambda'} \right|^2 \right\rangle \delta\left(\varepsilon_{\boldsymbol{k},\lambda} - \varepsilon_{\boldsymbol{k}',\lambda'}\right) \tag{H.11}$$

We solve Eq. (H.9) in the steady state and assuming that perturbations to the distribution function, in the linear response regime, are concentrated at the Fermi energy and are proportional to the applied thermal gradient:

$$f_{\boldsymbol{k},\lambda} = f^{(0)}_{\boldsymbol{k},\lambda} + \tau \left( -\frac{\partial f^{(0)}_{\boldsymbol{k},\lambda}}{\partial \varepsilon_{\boldsymbol{k},\lambda}} \right) \left( \frac{\varepsilon_{\boldsymbol{k},\lambda}}{T} - \frac{\partial \varepsilon_{\boldsymbol{k},\lambda}}{\partial T} \right) v_{\boldsymbol{k},\lambda} \cdot \nabla_{\boldsymbol{r}} T \tag{H.12}$$

In this equation, $\tau$ is the electron-impurity heat transport time, which is to be determined via the present calculation. By plugging Eq. (H.12) into Eq. (H.9), we obtain an equation that is solved via a standard projection on the heat-current mode. The method consists in multiplying such equation by the energy current transported by a single quasiparticle, $\varepsilon_{\boldsymbol{k},\lambda} v_{\boldsymbol{k},\lambda}$, and in summing it over all possible states, *i.e.* over $\boldsymbol{k}$ and $\lambda$. In this way, the integro-differential equation (H.9) reduces to an algebraic one, which can be solved to yield $\tau$. As a result, we are able to determine the heat transport time and the quasiparticle thermal conductivity. After few manipulations, the latter can be rewritten as:

$$\kappa = \frac{e^2}{\rho} \frac{D^2(\Delta(T), T)}{I(\Delta(T), T)} \tag{H.13}$$

where

$$D(\Delta(T), T) = \int_0^\infty d\xi \int \frac{d\varphi_{\boldsymbol{k}}}{2\pi} \left[ 4k_{\mathrm{B}} T \cosh^2\left( \frac{\sqrt{\xi^2 + \Delta^2_{\boldsymbol{k}}}}{2k_{\mathrm{B}} T} \right) \right]^{-1} \left( \frac{\xi^2 + \Delta^2_{\boldsymbol{k}}}{T} - \Delta_{\boldsymbol{k}} \frac{\partial \Delta_{\boldsymbol{k}}}{\partial T} \right) \tag{H.14}$$

$$I(\Delta(T), T) \simeq \frac{1}{2} \int_0^\infty d\xi \int \frac{d\varphi_{\boldsymbol{k}}}{2\pi} \left[ 4k_{\mathrm{B}} T \cosh^2\left( \frac{\sqrt{\xi^2 + \Delta^2_{\boldsymbol{k}}}}{2k_{\mathrm{B}} T} \right) \right]^{-1} \left( \frac{\xi^2 + \Delta^2_{\boldsymbol{k}}}{T} - \Delta_{\boldsymbol{k}} \frac{\partial \Delta_{\boldsymbol{k}}}{\partial T} \right) \frac{\xi}{\sqrt{\xi^2 + \Delta^2_{\boldsymbol{k}}}} \tag{H.15}$$

To obtain these equations, as usual in superconductivity, we have approximated:

$$N_{\mathrm{f}} \int \frac{d^2\boldsymbol{k}}{(2\pi)^2} \rightarrow N(\varepsilon_{\mathrm{F}}) \int_0^{\infty} d\xi \int \frac{d\varphi_{\boldsymbol{k}}}{2\pi} \tag{H.14}$$

where $N(\varepsilon_F)$ is the density of states at the Fermi energy of the normal system. The angle between the nodal line of the superconducting order parameter and the temperature gradient is unknown and may be different in different regions of the sample. Therefore, the quantities in Eq. (H.14) and (H.15) have been averaged over the directions of the applied temperature gradient (angles are measured from the direction of the nodal line for convenience). We have further approximated the expression of $I(\Delta(T), T)$ in Eq. (H.14) by using the fact that perfect backscattering is the dominant source of resistivity. Hence, we have set $\varphi_{\boldsymbol{k}'} = \varphi_{\boldsymbol{k}} + \pi$ , which in turn implies that $|\Delta_{\boldsymbol{k}}| = |\Delta_{\boldsymbol{k}'}|$. This, together with the δ-function in Eq. (H.11), constrained $|\xi_{\boldsymbol{k}}| = |\xi_{\boldsymbol{k}'}|$ and allowed us to perform the integral over $\boldsymbol{k}'$ analytically.